# A Study of the Early-stage Evolution of Relativistic Electron-Ion Shock using 3D PIC Simulations

E. J. Choi[1], K. Min[1], K.-I. Nishikawa[2] and C. R. Choi[1]

[1]*Korea Advanced Institute of Science and Technology, Daejeon 305-701, Republic of Korea*

[2]*Department of Physics, ZP12, University of Alabama in Huntsville, AL 35899, USA*

We report the results of a 3D particle-in-cell (PIC) simulation carried out to study the early-stage evolution of the shock formed when an unmagnetized relativistic jet interacts with an ambient electron-ion plasma. Full-shock structures associated with the interaction are observed in the ambient frame. When open boundaries are employed in the direction of the jet; the forward shock is seen as a hybrid structure consisting of an electrostatic shock combined with a double layer, while the reverse shock is seen as a double layer. The ambient ions show two distinct features across the forward shock: a population penetrating into the shocked region from the precursor region and an accelerated population escaping from the shocked region into the precursor region. This behavior is a signature of a combination of an electrostatic shock and a double layer. Jet electrons are seen to be electrostatically trapped between the forward and reverse shock structures showing a ring-like distribution in a phase-space plot, while ambient electrons are thermalized and become essentially isotropic in the shocked region. The magnetic energy density grows to a few percent of the jet kinetic energy density at both the forward and the reverse shock transition layers in a rather short time scale. We see little disturbance of the jet ions over this time scale.

Shocks in astrophysical objects such as supernova remnants (SNRs), active galactic nuclei (AGNs), and gamma-ray bursts (GRBs) are believed to accelerate particles efficiently, producing power-law energy spectra. Recent X-ray and gamma-ray observations indicate that electrons and protons may be accelerated to ~TeV energies in SNRs [1] [2] [3]. Afterglows in GRBs are attributed to inverse Compton scattering and to the synchrotron emission of high-energy electrons that are accelerated in the shocks produced the explosion of progenitor stars in their terminal stage [4] [5] [6]. Cosmic rays up to the knee in the cosmic ray spectrum (~$10^{15}$eV) are thought to be produced in diffusive SNR shocks [7] [8]. As evidenced by the synchrotron radiation from SNRs, shocks can also amplify magnetic fields [9]. However, the simple compressional amplification of magnetic fields by the shock may not account for the field strength, which needs to be high enough to effectively produce the observed synchrotron radiation. For example, the magnetic energy is believed to be comparable to the particle energy in GRB afterglows [10] [11]. Thus, the transverse filamentation instability, or Weibel instability, has been invoked as a viable mechanism for the efficient generation of strong magnetic fields [10].

The Weibel instability generates transverse electric and magnetic fields in plasmas with an anisotropic velocity distribution [12] [13] [14]. Electron current channels are formed by self-excited small-scale magnetic fields and are amplified by the two-stream-like filamentation instability [10] [15]. These electron current filaments merge with one another to form larger filaments, producing large-scale magnetic fields. Electron current filaments may accelerate the development of ion filamentation, although ion filaments can form even in the presence of a thermal plasma background [16]. The strong radial electric field, sustained by the magnetic pressure gradient force, accelerates electrons in the transverse direction [17]. The accelerated electrons show a power-law distribution of energies, which agrees with the observed synchrotron radiation spectra in the afterglow of GRBs. The process of field amplification and particle acceleration associated with the Weibel instability in collisionless relativistic shocks has been well demonstrated via particle-in-cell (PIC) simulations [18] ~ [29]. According to the simulations, filament formation is induced by the Weibel instability. Merging of the filaments was manifested by density and the magnetic field enhancement in the shock transition layers, and the magnetic energy grew to approximately 10 - 20 % of the jet kinetic energy. However, in the downstream region filaments broke up into turbulent clumps with much smaller field amplitudes accompanied by isotropization and thermalization of the particles.

In addition to the transverse electromagnetic fields mentioned above, it is well known that an electrostatic structure is generated in the longitudinal direction when two plasma beams collide [30] [31]. When two counter-streaming plasma clouds overlap, electrons with high mobility escape easily from the overlapped region, whereas ions do not, and this produces large ambipolar electric fields at the boundaries. Ions entering the overlapped region are slowed down in the boundary layers due to the higher potential in the overlapped region, while electrons are trapped. When the electric fields are strong enough to slow down the incoming ions to speeds comparable to the thermal velocities in the overlapped region, the electrostatic structures at the boundary layers develop into electrostatic shocks. Electrostatic shocks can generate shock-reflected ion beams, while ions in the downstream region can accelerate into the upstream region due to the same electric field, thus indicating the presence of double layers [32]. Such a hybrid structure consisting of double layers and electrostatic shocks are seen in recent PIC simulations [33] and in laser-driven plasma experiments [34]. We shall denote as shocks the plasma structures, across which the electron temperature and the ion speed is changing drastically. A shock in this definition can be an electrostatic shock in the stringent definition of Ref. [32], a double layer or a combination of both.

In this paper, we report on the results of a 3D PIC simulation which was performed to study the interaction between a relativistic jet and a stationary ambient medium. We note that most previous PIC simulations were carried out using a reflecting boundary to produce counter-streaming beams [26] [27] [28]. In these simulations, shocks were generated readily when the injected beams underwent head-on collisions with the reflected particle counter-streaming plasma. However, the results of these simulations do not show the full structure of the shocks as seen in the frame of the ambient medium. For example, in simulations using a reflecting boundary that represents a shock's contact discontinuity, it is not possible to distinguish between the leading and the trailing shocks that would appear on opposite sides of the shock contact discontinuity.

Thus, an open boundary condition in the flow direction is more useful for the study of the evolution of the full-shock structures that are generated by the collision of a jet with an ambient plasma. Nishikawa et al. (2009) [25] and Niemiec et al. (2013) [35] employed open boundary conditions in the flow direction to study shock development in electron-positron jet and ambient plasmas of different densities. They were able to observe full-shock structures consisting of a leading shock, a trailing shock, and a contact discontinuity. We have performed similar simulations for electron-ion plasmas using radiating open boundary conditions. We focus in this report on the early evolution of a shock system that is dominated by electron motions. The long-term evolution on the ion time scale was previously discussed in Niemiec et al. (2013) [35].

The present simulation was performed using a message passing interface (MPI)-based parallel version of the relativistic particle in-cell (RPIC) code TRISTAN [20] [25] [36]. The total size of the simulation box is $(L_x, L_y, L_z) = (8192\Delta, 64\Delta, 64\Delta)$. An unmagnetized electron-ion jet is initialized in the left region of the simulation box (x ≤ 512Δ), and flows in the positive x-direction with a velocity of $v_d = 0.998c$ ($\gamma = \sqrt{(1-(v_d/c)^2)} = 15$), where γ is the Lorentz factor and $c$ is the speed of light. The jet is continuously injected through the left boundary for the duration of the simulation. An unmagnetized electron-ion ambient plasma fills the right region of the simulation box (x ≥ 512Δ). The number densities of the jet and the ambient are both 8 particles per cell for both ions and electrons, with a total number of 200 million particles. The ion-to-electron mass ratio is $m_i/m_e = 16$, and the temperatures of the electrons and the ions are assumed to be identical. The thermal velocity of the ambient and jet electrons is 0.14c in their proper frames. The skin depths are 5Δ and 20Δ for the elections and ions, respectively, and the plasma frequencies are $\omega_{pe} = 0.2$ for the electrons and $\omega_{pi} = 0.05$ for the ions. Radiating open boundary conditions are imposed on the left (x = 0) and right (x = 8192Δ) boundaries, and periodic boundary conditions are used for the transverse boundaries [25]. Particles moving out through the left and the right open boundaries are assumed to be lost, while particles moving out through the periodic transverse boundaries are re-injected from the opposite boundary. The initial simulation setup is shown in Figure 1.

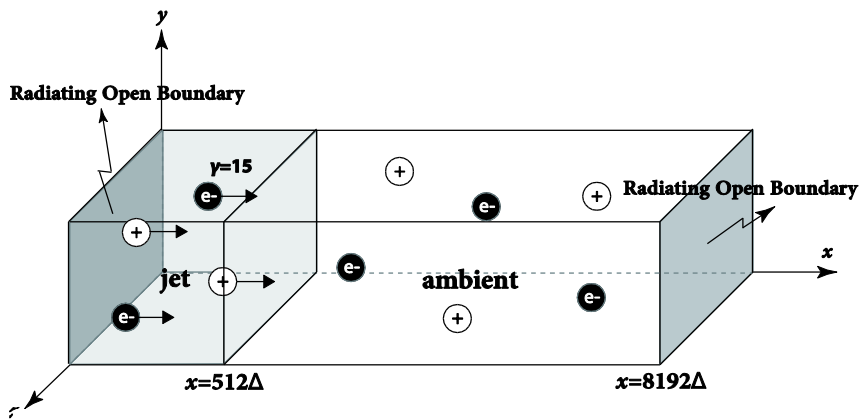

FIG. 1. Simulation setup: a jet located to the left of $x = 512\Delta$ moves in the +x-direction with a Lorentz factor $\gamma = 15$ and collides with an ambient plasma at rest that occupies the region beyond $x = 512\Delta$.

Figure 2 shows the evolution of the shock structure. The time-evolution of the total electron density profile and the peak electron density in the shocked region are shown in (a) and (b), respectively. The ion density (c), the electron density (d), three components of the magnetic field (e), and three components of the electric field (f) are plotted at the end of the simulation (T = $7372/\omega_{pe}$). All of the quantities in the figure are values averaged over the yz-plane. The simulation was terminated at T = $7372/\omega_{pe}$ when the jet reached the right boundary. Figure 2(b) shows that the shocked densities do not appear to change much after T = $6000/\omega_{pe}$ though the shock structure has not at this point reached a steady state, and the shock surfaces continue to become sharper.

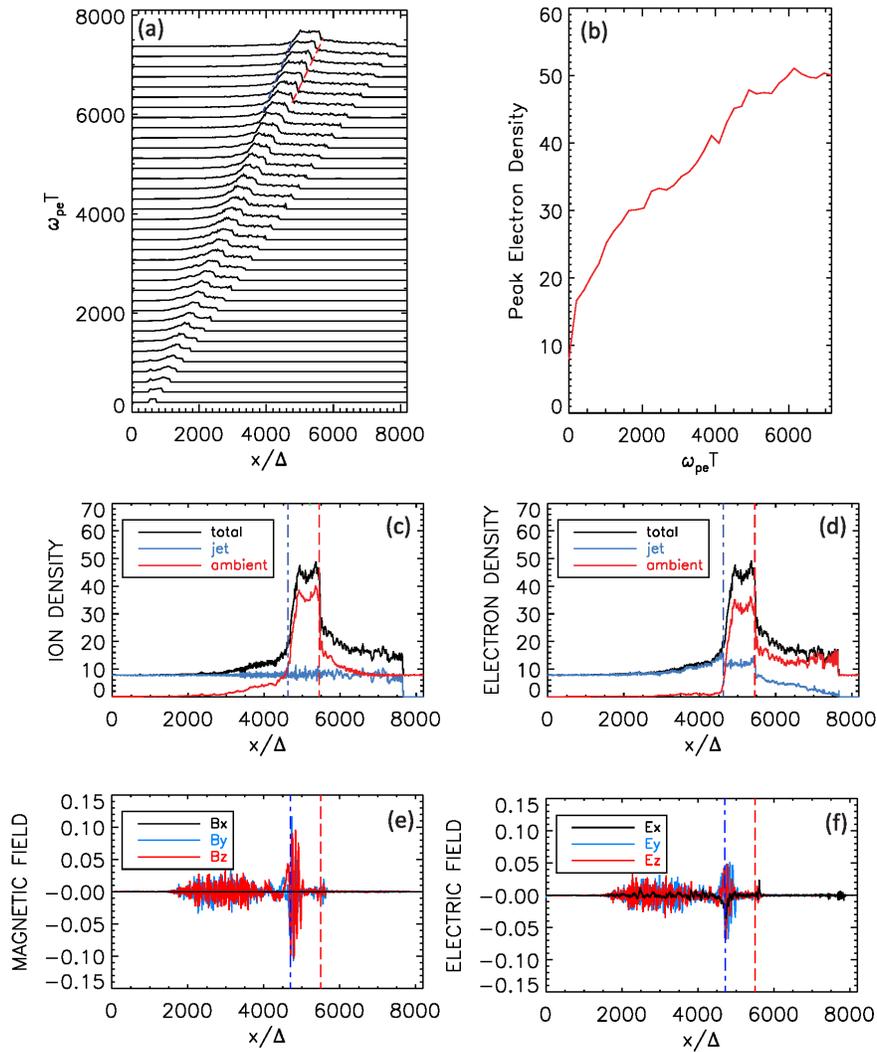

FIG. 2. Time evolution of the electron density profile and the peak electron density in the shocked region are shown in panels (a) and (b), respectively. Other panels show spatial profiles of (c) the ion density, (d) the electron density, (e) three components of the magnetic field, and (f) three components of the electric field at the end of the simulation ($\mathbf{T = 7372/\omega_{pe}}$). The electromagnetic fields are expressed in such a way that their energy densities are normalized by the jet kinetic energy density. All of the quantities in the figure are values averaged over the **yz**-plane.

First, we note that the present simulation results at the end of the simulation show a dual-shock structure that includes a forward (leading) shock at $x/\Delta \sim 5500$ and a reverse (trailing) shock at $x/\Delta \sim 4600$: the forward and reverse shocks are indicated by the red dashed and the blue dash-dotted lines, respectively. The estimated speeds of these structures, based on Figure 2(a), are $0.82c$ for the forward shock and $0.61c$ for the reverse shock. Figures 2(c) and 2(d) show the ion and the electron density profiles, respectively, at $T = 7372/\omega_{pe}$, and the ambient, the jet, and the total densities are shown in red, blue, and black, respectively. The figure shows that the ambient density increases behind the forward shock at $x/\Delta \sim 5500$ by a factor of five for both the ions and the electrons. However, the jet ion density does not show a significant change across the forward shock, while we see a sharp increase in the jet electron density behind the forward shock. In the precursor region in front of the forward shock, the jet electron density decreases gradually in front of the shock while the jet ion density does not change much in front of the shock all the way to the right boundary of the simulation box. It is interesting to note that the ambient medium dominates the reverse shock structure at $x/\Delta \sim 4600$. Behind the reverse shock, both the ambient ion and ambient electron densities decrease sharply with a region of low ambient density moving with the reverse shock. The jet electron density shows a slight enhancement at the reverse shock, but the jet ion density does not change across the reverse shock.

The electromagnetic fields associated with the shock are shown in Figures 2(e) and 2(f), where the electromagnetic fields are expressed in such a way that their energy densities are normalized by the jet kinetic energy density. We note that the transverse components of the magnetic fields are enhanced significantly and peak in the forward and reverse shock regions. The magnetic field is weaker in the forward shock region than in the reverse shock region but is in general comparable. We also note that the transverse components of the electric field are enhanced at the forward and reverse shocks. What is interesting, however, is the enhancement of the longitudinal component of the electric field at both the forward and reverse shocks. These ambipolar electric fields are caused by the ion density gradient and the different mobilities of electrons and ions [30] [31]. We find that the shocks are hybrid structures consisting of electrostatic shocks and double layers, as is indicated by the phase-space plots shown in Figure 3.

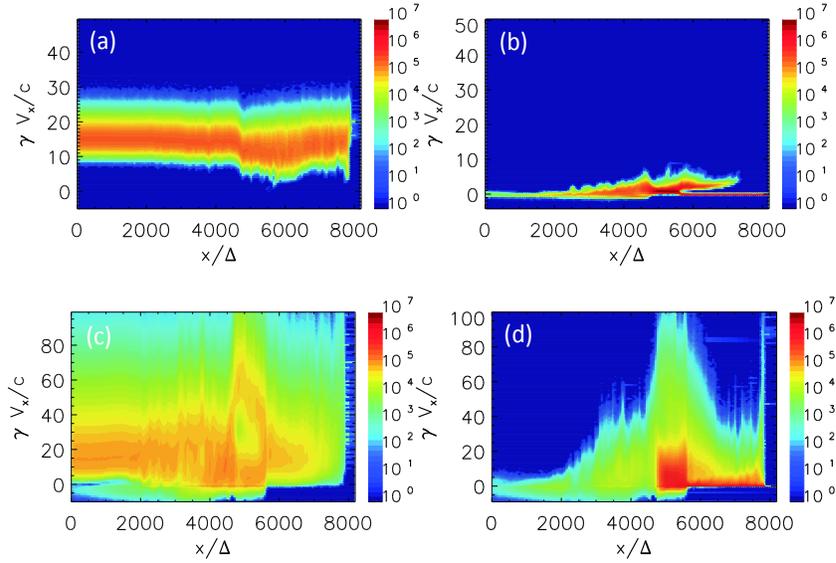

FIG. 3. Plot of $\gamma v_x/c$ as a function of $x$ for (a) the jet ions, (b) the ambient ions, (c) the jet electrons, and (d) the ambient electrons at $T = 7372/\omega_{pe}$. The color bars indicate the total number of particles in the unit interval $\Delta$ in the x-direction, summed over the yz-plane.

Figure 3 shows the velocity $\gamma v_x/c$ of both the jet (panels a and c) and ambient (panels b and d) particles along the x direction. We can easily identify the shocked region between x/Δ ~4600 and x/Δ ~5500, especially from the plot of the ambient electrons shown in panel (d). In the precursor region in front of the forward shock, see Figure 3(a), jet ions interact with ambient particles and are slowed down gradually by ~20 % from a leading edge Lorentz factor of γ = 15 to a minimum value of γ~12 in front of the forward shock. At the reverse shock the jet ions experience a sharp decline in Lorentz factor from the original value of γ = 15. Figure 3(b) shows that the ambient ions have relatively low velocities in the region between the reverse shock at x/Δ ~4600 and the forward shock at x/Δ ~5500. Ambient ions originally located in front of the forward shock, penetrate the forward shock and are accelerated by the electrostatic field at the shock, a natural consequence of the potential structure associated with an electrostatic shock. It is also remarkable that there are fast ambient ions moving in front of the forward shock. These ambient ions represent a population escaping from a shocked region of higher potential through the forward shock to a region of lower potential, and this behavior is the signature of a double layer. Hence, the forward shock observed in the present simulation is actually a hybrid structure consisting of an electrostatic shock combined with a double layer [32]. We also note a population of heated ambient ions in the region trailing the reverse shock. This population is also the signature of a double layer as the ions move from a higher potential region to a region of lower potential.

Electrons also undergo peculiar changes, as shown in Figures 3(c) and 3(d). First, we note that the jet electrons show a ring-like phase-space structure in the shocked region. The jet electrons are accelerated across the reverse shock into a shocked region of higher potential, but cannot penetrate the forward shock to a region of lower potential, thereby becoming trapped in the shocked region. The ambient electrons also show strong trapping in the shocked region. While most of the ambient electrons become thermalized, some of them are accelerated above $\gamma v_x = 80$. Both jet and ambient electrons are sound to be heated in the region trailing the reverse shock. This is most likely caused by the fluctuating fields that trail the reverse shock, see Figure 2.

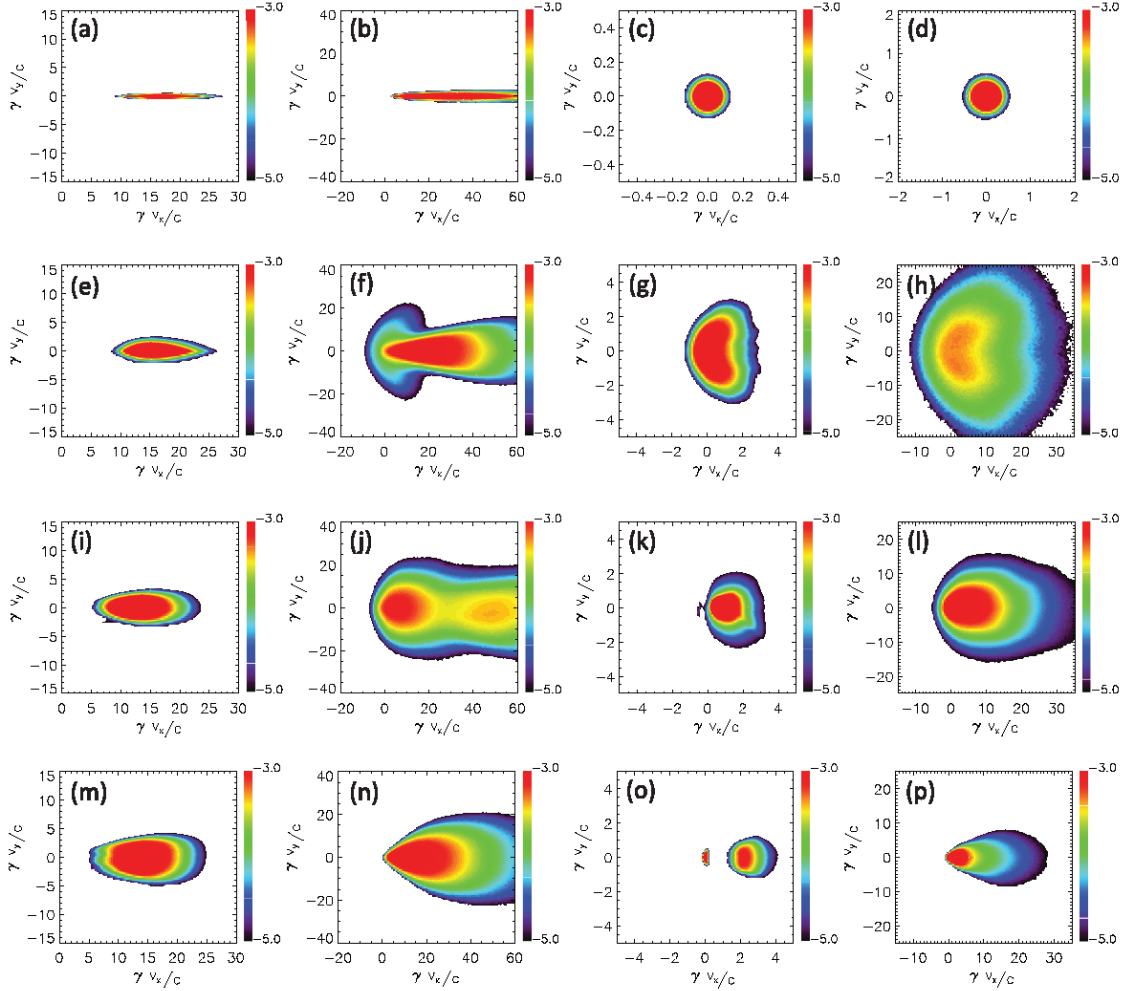

FIG. 4. Phase-space plots of $\gamma v_x/c$ - $\gamma v_y/c$ for jet and ambient particles where panels in the top row show the initial setup and columns show jet ions, jet electrons, ambient ions, and ambient electrons, from left to right. The color bars give the number density log scale. All other panels are at $T = 7372/\omega_{pe}$. Panels in the second row are from the trailing scattering region between $x/\Delta = 2000$ and $x/\Delta = 4000$. Panels in the third row correspond to the shocked region between $x/\Delta = 4600$ and $x/\Delta = 5500$. Panels in the fourth row correspond to the forward scattering region between $x/\Delta = 6000$ and $x/\Delta = 7500$.

Figure 4 shows phase-space plots in $\gamma v_x/c$ - $\gamma v_y/c$ for the jet and ambient particles at initial setup (top row) and at $T = 7372/\omega_{pe}$ in the trailing scattering region between $x/\Delta = 2000$ and $x/\Delta = 4000$ (second row), the downstream region between $x/\Delta = 4600$ and $x/\Delta = 5500$ (the third row), and the precursor region between $x/\Delta = 6000$ and $x/\Delta = 7500$ (fourth row). The columns correspond to jet ions, jet electrons, ambient ions, and ambient electrons, from left to right. First, we note that jet ions undergo gradual transverse heating as they propagate through the trailing scattering region (panel e), the shocked region (panel i), and the precursor region (panel m), but only minimal deceleration is noted in the longitudinal direction in the shocked (panel i) and the precursor (panel m) regions. Hence, the present result corresponds to evolution on a time scale shorter than that of an ion beam interaction. The jet electrons, on the other hand, show significant transverse heating even in the trailing scattering region (panel f). The two distinct populations of faster moving and slower moving electrons in the shocked region (panel j) correspond to the ring-like distribution in the phase-space plot in Figure 3(c). It is remarkable that each of the two populations is well thermalized. The jet electrons are also well thermalized in the precursor region (panel n). The ambient ions and electrons are heated and well thermalized in the shocked and the trailing scattering regions, though transverse heating appears more significant than longitudinal heating in the trailing scattering region. The accelerated ambient ions identified in Figure 3(b) appear here at $\gamma v_x \sim 2$ in panel (o) that shows the precursor region in front of the forward shock. These ions also appear to be heated in both the longitudinal and the transverse directions to become relatively isotropic around the drift speed. On the other hand, panel (p) shows that the ambient electrons experience more longitudinal heating than transverse heating in the precursor region.

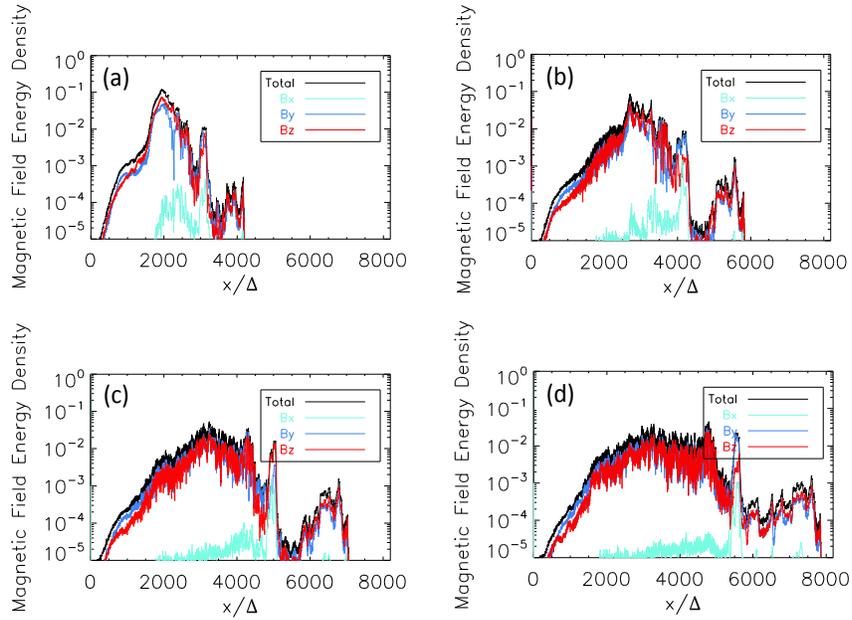

FIG. 5. Spatial profiles of the energy densities corresponding to the three magnetic field components (a) at $T = 3686/\omega_{pe}$, (b) at $T = 5324/\omega_{pe}$, (c) at $T = 6553/\omega_{pe}$, and (d) at $T = 7372/\omega_{pe}$. Energy densities are averaged over the yz-plane and divided by the jet energy density.

Spatial profiles of the energy densities corresponding to the three magnetic field components are shown in Figure 5 at times (a) T = $3686/\omega_{pe}$, (b) T = $5324/\omega_{pe}$, (c) T = $6553/\omega_{pe}$, and (d) T = $7372/\omega_{pe}$. Magnetic energy densities are averaged over the yz-plane and divided by the jet energy density. The figures show that the field energies are dominated by the transverse components. We first note that strong magnetic field energy is generated in the trailing scattering region. This is expected as the Weibel instability should be prominent in the region where the high speed jet directly interacts with the ambient medium. Figure 5(a) shows the magnetic energy density peak in the trailing scattering region at T = $3686/\omega_{pe}$ at ~ 10% of the jet kinetic energy density. This value is comparable to the level of the magnetic energy density noted in previous longer duration simulations in which the jet ions contribute to the Weibel instability. While the magnetic field energy density decreases with time, it still remains at a few percent of the jet kinetic energy density at the end of the simulation, e.g., Figure 5(d). The magnetic energy densities in the forward and the reverse shocks are both a few percent of the jet kinetic energy density, whereas the magnetic field energy density in the shocked region is rather low, at a level of ~ 0.1% of the jet kinetic energy density.

In summary, we performed 3D PIC simulations to study the early stage characteristics of shock evolution associated with an unmagnetized relativistic jet injected into unmagnetized electron-ion plasma. The open boundaries in the jet direction enabled us to examine the formation of forward and reverse shock structures in the ambient frame. It was found that the forward shock is a hybrid of an electrostatic shock and a double layer while the reverse shock shows the characteristics of a double layer. The roles of forward and reverse shocks will switch if the system is observed in a frame moving with the jet. While the evolution of the shock structures seen in the jet frame will be slightly different since both the jet and the ambient densities change, we still expect to observe similar structures due to symmetry. Hence, the features observed in the present simulation could actually represent four shock structures attached to the boundaries of the shocked region. In this simulation the magnetic energy density was observed to be enhanced at both the forward and the reverse shock transition layers with values of a few percent of the jet kinetic energy density on a rather short time scale. In the shocked region between forward and reverse shocks the magnetic energy density is only ~ 0.1% of the jet kinetic energy density. These results are in agreement with previous simulations on a much longer time scale using reflecting boundaries [28] [29].

**ACKNOWLEDGMENTS**

We thank the referee for insightful comments and suggestions that led correct interpretation of the shocks observed in the present simulation. We acknowledge useful comments and discussion with Phil Hardee and Jacek Niemiec that helped to improve this report. K.-W. Min acknowledges the support by the National Research Foundation of Korea through its grant

NRF-2013M1A3A3A02041911. K.I. is supported by NSF AST-0908040, NASA NNX08AG83G, NNX08AL39G, NNX09AD16G, NNX12AH06G, NNX13AP14G, and NNX13AP21G. The simulations were performed on the Columbia and Pleiades supercomputers at the NASA Advanced Supercomputing (NAS) center, and on Kraken at NICS and Ranger at TACC with the XSEDE project, which is supported by the NSF. Part of this work was initiated while K.-I. N. was visiting KAIST.